\documentclass[12]{article}
\usepackage{amsmath}
\usepackage{graphicx}
\begin{document}

\Large
\begin{center}
\bf{A Combinatorial Grassmannian Representation of the\\ Magic Three-Qubit Veldkamp Line}
\end{center}
\vspace*{-.3cm}
\begin{center}
Metod Saniga
\end{center}
\vspace*{-.5cm} \normalsize
\begin{center}
Astronomical Institute, Slovak Academy of Sciences\\
SK-05960 Tatransk\' a Lomnica, Slovak Republic\\
(msaniga@astro.sk)  
\end{center}

\vspace*{-.4cm} \noindent \hrulefill

\vspace*{-.1cm} \noindent {\bf Abstract}

\noindent
It is demonstrated that the magic three-qubit Veldkamp line occurs naturally within the Veldkamp space of combinatorial Grassmannian of type $G_2(7)$, $\mathcal{V}(G_2(7))$. The lines of the ambient symplectic polar space are those lines of $\mathcal{V}(G_2(7))$ whose cores feature an odd number of points of $G_2(7)$. After introducing basic properties of three different types of points and six distinct types of lines of $\mathcal{V}(G_2(7))$, we explicitly show the combinatorial Grassmannian composition of the magic Veldkamp line; we first give representatives of points and lines of its core generalized quadrangle GQ$(2,2)$,  and then additional points and lines of a specific elliptic quadric $\mathcal{Q}^{-}$(5,\,2), a hyperbolic quadric $\mathcal{Q}^{+}$(5,\,2) and a quadratic cone $\widehat{\mathcal{Q}}$(4,\,2) that are centered on the GQ$(2,2)$.  In particular, each point of $\mathcal{Q}^{+}$(5,\,2) is represented by a Pasch configuration and its complementary line, the (Schl\"afli) double-six of points in $\mathcal{Q}^{-}$(5,\,2) comprise six Cayley-Salmon configurations and six Desargues configurations with their complementary points, and the remaining  Cayley-Salmon configuration stands for the vertex of $\widehat{\mathcal{Q}}$(4,\,2).\\

\vspace*{-.2cm}


\vspace*{-.2cm} \noindent \hrulefill

\section{Introduction}
One of the most startling results of finite-geometric approach to the field of quantum information and the so-called black-hole/qubit correspondence is, undoubtedly, a recent discovery \cite{lsz,lhs} of the existence of a magic Veldkamp line associated with the five-dimensional binary symplectic polar space underlying geometry of the three-qubit Pauli group. Three basic constituents of this line (illustrated\footnote{Symbols and notation are explained in the next section.} graphically in Figure 1) host a number of extensions of generalized quadrangles with lines of size three isomorphic to affine polar spaces of rank three and order two, each having distinguished physical interpretation and in their totality offering a remarkable unifying framework for form theories of gravity and black hole entropy. The purpose of this paper is to show that this magic line has also a remarkable representation in the Veldkamp space of combinatorial Grassmannian of type $G_2(7)$.  

\begin{figure}[pth!]
	\centering
	\includegraphics[width=9truecm]{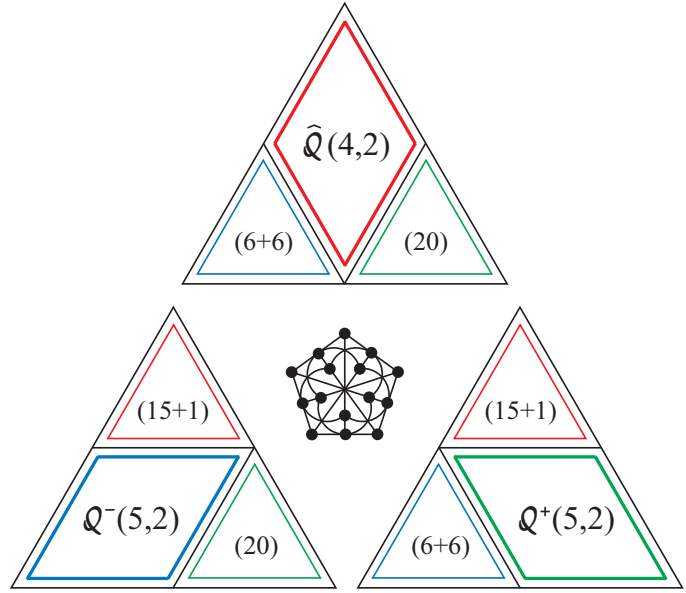}
	\caption{A sketch of the structure of the magic three-qubit Veldkamp line comprising an elliptic quadric ($\mathcal{Q}^{-}$(5,\,2) $\cong$ GQ(2,\,4); represented by a blue rhombus), a hyperbolic quadric 
	($\mathcal{Q}^{+}$(5,\,2); green rhombus) and a 
	quadratic cone ($\widehat{\mathcal{Q}}$(4,\,2); red rhombus), the three objects having a $\mathcal{Q}$(4,\,2) $\cong$ GQ(2,\,2) in common (illustrated by a black `doily' in the middle). The numbers inside triangles indicate the number of points in the complement of GQ(2,\,2) of the geometrical object in question.}
	\label{fig1}
\end{figure}

\section{Relevant Finite-Geometrical Background}
To this end, we first give an overview of relevant finite geometry.
We start with a finite {\it point-line incidence structure} $\mathcal{C} = (\mathcal{P},\mathcal{L},I)$ where $\mathcal{P}$ and $\mathcal{L}$ are, respectively, finite sets of points and lines and where incidence $I \subseteq \mathcal{P} \times \mathcal{L}$ is a binary relation indicating which point-line pairs are incident (see, e.\,g., \cite{shult}).    
Here, we shall only be concerned with specific point-line incidence structures called {\it configurations} \cite{grun}. A $(v_r,b_k)$-configuration is a $\mathcal{C}$ where: 1) $v = \vert \mathcal{P} \vert$ and $b = \vert \mathcal{L}\vert$, 2) every line has $k$ points and every point is on $r$ lines, and 3) two distinct lines intersect in at most one point and every two distinct points are joined by at most one line; a configuration where $v=b$ and $r=k$ is called symmetric (or balanced), and usually denoted as a $(v_r)$-configuration.
A $(v_r,b_k)$-configuration with $v = {r+k-1 \choose r}$ and $b = {r+k-1 \choose k}$ is called a $binomial$ configuration.
Next, a {\it geometric hyperplane} of $\mathcal{C} = (\mathcal{P},\mathcal{L},I)$ is a proper subset of $\mathcal{P}$ such that a line from $\mathcal{C}$ either lies fully in the subset, or shares with it only one point. 
 If $\mathcal{C}$ possesses geometric hyperplanes, then one can define the {\it Veldkamp space} of $\mathcal{C}$,  $\mathcal{V}(\mathcal{C})$, as follows \cite{buec}: (i) a point of $\mathcal{V}(\mathcal{C})$ is a geometric hyperplane of  $\mathcal{C}$
and (ii) a line of $\mathcal{V}(\mathcal{C})$ is the collection $H'H''$ of all geometric hyperplanes $H$ of $\mathcal{C}$  such that $H' \cap H'' = H' \cap H = H'' \cap H$ or $H = H', H''$, where $H'$ and $H''$ are distinct geometric hyperplanes; the set $H' \cap H''$ is sometimes called the core. If each line of $\mathcal{C}$ has three points a line of $\mathcal{V}(\mathcal{C})$ is also of size three and of the form $\{H', H'', \overline{H' \Delta H''}\}$, where the symbol $\Delta$ stands for the symmetric difference of the two geometric hyperplanes and an overbar denotes the complement of the object indicated. 
Our central concept is that of a combinatorial Grassmannian (see, e.\,g., \cite{pra1,pra2}) $G_k(|X|)$, where $k$ is a positive
integer and $X$ is a finite set, which is a point-line incidence structure whose points are $k$-element subsets of $X$ and whose lines are $(k + 1)$-element subsets of $X$, incidence being inclusion. It is known \cite{pra1} that if $|X| =
N$ and $k=2$,
$G_2(N)$ is a binomial $\left({N \choose 2}_{N-2}, {N \choose
3}_{3}\right)$-configuration; in particular, $G_2(3)$ is a single line,  $G_2(4)$ is the Pasch $(6_2,4_3)$-configuration, 
$G_2(5)$ is the Desargues $(10_3)$-configuration and $G_2(6)$ is the Cayley-Salmon $(15_4, 20_3)$-configuration \cite{saetal}.

A (finite-dimensional) classical polar space (see, for example, \cite{ht,cam}) describes the geometry of a $d$-dimensional vector space
over the Galois field GF$(q)$, $V(d, q)$, carrying a non-degenerate reflexive
sesquilinear form $\sigma(x, y)$. The polar space is called symplectic,
and usually denoted as $\mathcal{W}(d-1,q)$,  if this form is bilinear
and alternating, i.e., if $\sigma(x, x) = 0$ for all $x \in V(d,
q)$; such a space exists only if $d=2N$, where $N \geq 2$ is called its
rank. A subspace of $V(d, q)$ is called totally isotropic if
$\sigma$ vanishes identically on it. $\mathcal{W}(2N-1,q)$ can then be
regarded as the space of totally isotropic subspaces of the ambient space PG$(2N-1,
q)$, the ordinary $(2N - 1)$-dimensional projective space over
GF$(q)$, with respect to a symplectic form (also known as a null
polarity).
A quadric in PG$(d, q)$, $d \geq 1$, is the set of points whose coordinates satisfy an equation of the form $\sum_{i,j=1}^{d+1} a_{ij} x_i x_j = 0$, where at least one $a_{ij} \neq 0$.
Up to transformations of coordinates, there is one or two distinct kinds of non-singular quadrics in PG$(d, q)$ according as $d$ is even or odd, namely \cite{ht}:
 $\mathcal{Q}(2N,q)$, the {\it parabolic} quadric formed by all points of PG$(2N, q)$ satisfying the standard equation $x_1x_2+\cdots+x_{2N-1}x_{2N} + x_{2N+1}^{2} = 0$;
 $\mathcal{Q}^{-}(2N - 1,q)$, the {\it elliptic} quadric formed by all points of PG$(2N - 1, q)$ satisfying the standard equation $f(x_1,x_2)+x_3x_4+\cdots+x_{2N-1}x_{2N} = 0$, where $f$ is irreducible over GF$(q)$; and
 $\mathcal{Q}^{+}(2N - 1,q)$, the {\it hyperbolic} quadric formed by all points of PG$(2N - 1, q)$ satisfying the standard equation $x_1x_2+x_3x_4+\cdots+x_{2N-1}x_{2N} = 0$,
where $N \geq 1$. 
The number of points lying on quadrics is as follows \cite{ht}:
 $|Q(2N,q)|_p = (q^{2N}-1)/(q-1)$,
 $|Q^{-}(2N - 1,q)|_p = (q^{N-1}-1)(q^{N}+1)/(q-1) $,
 $|Q^{+}(2N - 1,q)|_p = (q^{N-1}+1)(q^{N}-1)/(q-1)$.
Given the hyperbolic quadric $\mathcal{Q}^{+}(2N - 1, q)$ of PG$(2N - 1, q)$, $N \geq 2$, a set $S$ of points such that each line joining two distinct points of $S$ has no point in common with $\mathcal{Q}^{+}(2N - 1, q)$  is called an exterior set of the quadric. It is known  that $|S| \leq (q^N - 1)/(q - 1)$; if $|S| = (q^N - 1)/(q - 1)$, then $S$ is called a maximal exterior set.  Interestingly \cite{thas}, $\mathcal{Q}^{+}(5, 2)$ has, up to isomorphism, a {\it unique} such set --- also known, after its discoverer, as a {\it Conwell hetpad} \cite{con}. 

Finally, one has to introduce a finite {\it generalized quadrangle} of order $(s, t)$, usually denoted GQ($s, t$), which is a $\mathcal{C}$ satisfying the following axioms \cite{paythas}: (i) each point is incident with $1 + t$ lines ($t \geq 1$) and two distinct points are incident with at most one line; (ii) each line is incident with $1 + s$ points ($s \geq 1$) and two distinct lines are incident with at most one point;  and (iii) if $x$ is a point and $L$ is a line not incident with $x$, then there exists a unique line through $x$ that is incident with $L$; from these axioms it readily
follows that $|\mathcal{P}| = (s+1)(st+1)$ and $|\mathcal{L}| = (t+1)(st+1)$. In what follows we shall only be concerned with its two particular types: GQ$(2,2) \cong \mathcal{Q}(4,2) \cong \mathcal{W}(3,2)$ and
GQ$(2, 4) \cong \mathcal{Q}^{-}(5,\,2)$.

\section{Veldkamp Space of $\boldsymbol{G_2(7)}$}
The Veldkamp space of $G_2(7)$, $\mathcal{V}(G_2(7))$, is isomorphic to PG$(5,2)$ and its was analyzed in detail in \cite{saetal} from which we highlight its basic properties. Let us take $X = \{1,2,3,4,5,6,7\}$ and assume that $a,b,c,d,e,f,$ and $g$ -- all different -- belong to $X$. The 63 points of $\mathcal{V}(G_2(7))$ are of three different types, as shown in Table 1, whereas its 651 lines fall into six distinct orbits, whose properties are given in Table 2; here, for example,  $abcd$:$efg$ indicates both a particular partition of $X$ into two complementary sets (i.\,e., $\{a,b,c,d\}$ and $\{e,f,g\}$) and the two combinatorial Grassmannians defined on these sets (i.\,e., $G_2(4)$ and $G_2(3)$).  We briefly note that every point of $\mathcal{V}(G_2(7))$ is a pair of complementary Grassmannians, and that there are no lines of type $(\alpha, \gamma, \gamma)$, $(\beta, \beta, \gamma)$ and $(\gamma, \gamma, \gamma)$.

\begin{table}[pth!]
\begin{center}
\caption{The three different types of points of $\mathcal{V}(G_2(7))$. } 
\begin{tabular}{||c|l|l|r||}
\hline \hline
Type      &  Form                 &  Geometrical constituents                                &  Number \\
\hline
$\alpha$  &  $abcd$:$efg$         &  Pasch configuration and its complementary line          &  35  \\
$\beta$   &  $abcde$:$fg$         &  Desargues configuration and its complementary point     &  21  \\
$\gamma$  &  $abcdef$:$g$         &  Cayley-Salmon configuration                             &  7  \\
 \hline \hline
\end{tabular}
\end{center}
\end{table}

\begin{table}[pth!]
\begin{center}
\caption{The seven different types of lines of $\mathcal{V}(G_2(7))$. } 
\begin{tabular}{||c|l|l|r||}
\hline \hline
Type                        &  Form                  & Core composition                                       &  Number \\
\hline
                            &  $abcd$:$efg$          &                                                        &                          \\
$(\alpha, \alpha, \alpha)$  &  $abef$:$cdg$          &  three mutually non-collinear points ($ab$,$cd$, and $ef$) &  105 \\
														&  $cdef$:$abg$          &                                                        &                          \\
\hline
                            &  $abcd$:$efg$          &                                                        &                          \\
$(\alpha, \alpha, \beta)$   &  $abce$:$dfg$          &  a line ($abc$) and a point ($fg$)                         &  210                     \\
														&  $abcfg$:$de$          &                                                        &                       \\
\hline
                            &  $abc$:$defg$          &                                                        &                       \\
$(\alpha, \alpha, \gamma)$  &  $def$:$abcg$          &  two disjoint lines ($abc$ and $def$)                  &  70                    \\
														&  $abcdef$:$g$          &                                                        &                       \\
\hline
                            &  $abcd$:$efg$          &                                                        &                        \\
$(\alpha, \beta, \beta)$    &  $ab$:$cdefg$          &  a line ($efg$) and two non-collinear points ($ab$ and $cd$)   &     105           \\
														&  $cd$:$abefg$          &                                                        &                       \\
\hline
                            &  $abcd$:$efg$          &                                                        &                          \\
$(\alpha, \beta, \gamma)$   &  $abcde$:$fg$          &  a Pasch configuration ($abcd$) and a point ($fg$)         &   105                    \\
														&  $abcdfg$:$e$          &                                                        &                       \\
\hline
                            &  $abcde$:$fg$          &                                                        &                        \\
$(\beta, \beta, \beta)$     &  $abcdf$:$eg$          &  a Pasch configuration  ($abcd$)                         &  35                    \\
														&  $abcdg$:$ef$          &                                                        &                       \\
\hline
                            &  $abcde$:$fg$          &                                                        &                        \\
$(\beta, \gamma, \gamma)$   &  $abcdef$:$g$          &  a Desargues  configuration   ($abcde$)                  &  21                       \\
														&  $abcdeg$:$f$          &                                                        &                       \\
\hline \hline
\end{tabular}
\end{center}
\end{table}

\noindent
In $\mathcal{V}(G_2(7))$, there exists a distinguished symplectic polar space $\overline{\mathcal{W}}$(5,\,2) whose lines comprise three orbits of lines of type $(\alpha, \alpha, \alpha)$,  $(\alpha, \beta, \beta)$ and $(\alpha, \beta, \gamma)$ --- that is the orbits whose cores feature an {\it odd} number of points of $G_2(7)$. Another prominent geometrical objects of $\mathcal{V}(G_2(7))$ are:
 a hyperbolic quadric $\mathcal{Q}^{+}_{0}$(5,\,2) $\in$ $\overline{\mathcal{W}}$(5,\,2) formed by 35 points of type $\alpha$ and 105 lines of type $(\alpha, \alpha, \alpha)$;  
 a combinatorial Grassmannian $G_2(7)$ formed 21 points of type $\beta$ and 35 lines of type $(\beta, \beta, \beta)$; 
and a Conwell heptad with respect to the above-defined  $\mathcal{Q}^{+}_{0}$(5,\,2) represented by seven points of type $\gamma$ (see also \cite{san}).

\section{Magic Three-Qubit Veldkamp Line in $\boldsymbol{\mathcal{V}(G_2(7))}$}
There are seven distinguished magic Veldkamp lines living in $\overline{\mathcal{W}}$(5,\,2), one per each element of $X$. A representative of them, also depicted in Figure 2, is structured as follows: 

\begin{itemize} 
\item Core GQ(2,\,2): Its 15 points are represented by those $\alpha$-points that share one digit in the second set, that is by points whose representatives are $abcd$:$ef7$ if the common digit is `7'; its 15 lines are those of type $(\alpha, \alpha, \alpha)$ of the following particular form

\begin{center}
\begin{tabular}{l}
$abcd$:$ef7$, \\
$abef$:$cd7$, \\
$cdef$:$ab7$.
\end{tabular}
\end{center}

\item $\mathcal{Q}^{-}$(5,\,2) $\cong$ GQ(2,\,4): The 12 additional points (the double-six) are represented by six $\beta$-points of the form $abcde$:$f7$  and  six $\gamma$-points of the form  $abcde7$:$f$; the 30 additional lines lie in the  $(\alpha, \beta, \gamma)$-orbit, being of the (complementary) form

\begin{center}
\begin{tabular}{ll}
$abcd$:$ef7$, & $abcd$:$ef7$, \\
$abcdf$:$e7$, & $abcde$:$f7$, \\
$abcde7$:$f$, & $abcdf7$:$e$.
\end{tabular}
\end{center}

\item $\mathcal{Q}^{+}$(5,\,2) $\equiv$ $\mathcal{Q}^{+}_{0}$(5,\,2): The 20 additional points are represented by $\alpha$-points of the form $abc7$:$def$; the 90 additional lines, belonging to the $(\alpha, \alpha, \alpha)$-orbit, read

\begin{center}
\begin{tabular}{ll}
$abcd$:$ef7$, & $abcd$:$ef7$,\\
$abe7$:$cdf$, & $abf7$:$cde$,\\
$cde7$:$abf$, & $cdf7$:$abe$.
\end{tabular}
\end{center}

\item $\widehat{\mathcal{Q}}$(4,\,2): The 16 additional points are represented by 15 $\beta$-points of the form $abcd7$:$ef$ and a single $\gamma$-point $abcdef$:$7$ (the vertex of the cone); the 15 additional lines are located in the $(\alpha, \beta, \gamma)$-orbit, having the form

\begin{center}
\begin{tabular}{l}
$abcd$:$ef7$, \\
$abcd7$:$ef$, \\
$abcdef$:$7$.
\end{tabular}
\end{center}

\end{itemize}

\begin{figure}[t]
	\centering
	\includegraphics[width=9truecm]{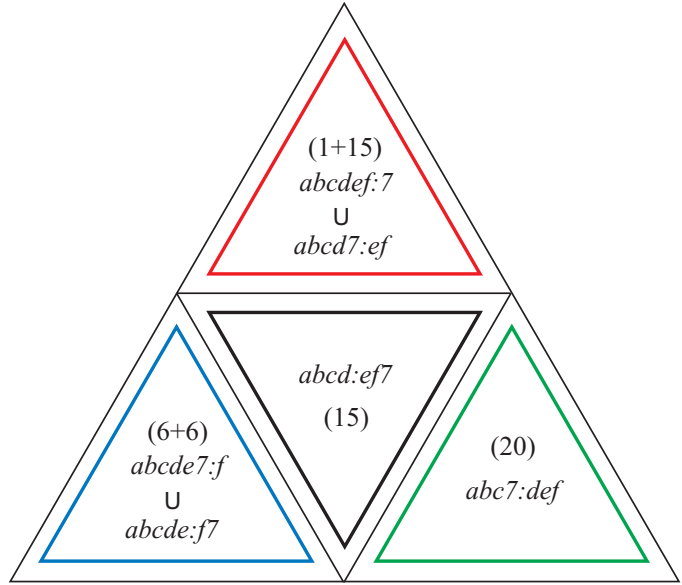}
	\caption{A pictorial representation of the four sectors of the magic Veldkamp line by different types of points of $\overline{\mathcal{W}}$(5,\,2)  $ \in \mathcal{V}(G_2(7))$. }
	\label{fig2}
\end{figure}

\section{Conclusion}
We have demonstrated that the Veldkamp space $\mathcal{V}(G_2(7))$  provides a rather natural environment for the magic Veldkamp line of three-qubits.  Interestingly, $\mathcal{V}(G_2(7))$ was recently found to be also related to finite geometry behind the 64-dimensional real Cayley-Dickson algebra \cite{saetal}. Hence, our findings seem to indicate that the nature of magic Veldkamp line may well have something to do with this particular algebra.

\section*{Acknowledgment}
This work was supported by the VEGA Grant Agency, Project 2/0003/16, as well as by the Austrian Science Fund (Fonds zur F\"orderung der Wissenschaftlichen Forschung (FWF)), Research Project M1564--N27. I am grateful to my friend Petr Pracna for electronic versions of the figures.

\end{document}